\documentclass[12 pt]{article}
\usepackage[utf8]{inputenc}
\usepackage{authblk}
\usepackage{epigraph}
\usepackage{hyperref}
\usepackage{graphicx}
\usepackage{mathtools}   
\usepackage{amsfonts}
\usepackage{appendix}
\usepackage{caption}
\usepackage{orcidlink}
\graphicspath{ {./figures/} }
\hypersetup{
    colorlinks=true,
    linkcolor=blue,
    urlcolor=blue,
}

\usepackage[style=apa,url=true,sorting=nyt,natbib=true]{biblatex}
\providecommand{\keywords}[1]
{
  \small	
  \textbf{\textit{Keywords---}} #1
}

\begin{document}

\title{{Against Radical Relationalism: in Defense of the Ordinal Structure of Time\footnote{Cite as: Mozota Frauca, Á. Against Radical Relationalism: in Defense of the Ordinal Structure of Time. \textit{Found Phys} 55, 37 (2025). \url{https://doi.org/10.1007/s10701-025-00850-5}}}}

\author[1,2]{Álvaro Mozota Frauca}
\affil[ ]{alvaro.mozota@upc.edu, \orcidlink{0000-0002-7715-0563} \href{https://orcid.org/0000-0002-7715-0563}{https://orcid.org/
0000-0002-7715-0563}}
\affil[1]{Department of Mathematics, Universitat Politècnica de Catalunya, Pau Gargallo 14, 08028 Barcelona, Spain}
\affil[2]{Department of Condensed Matter Physics, Universitat de Barcelona, Martí i Franquès 1, 08028 Barcelona, Spain}


%
%
%
%
%


\maketitle

\begin{abstract}
Some authors in the quantum gravity community endorse, explicitly or implicitly, a radical relationalist view of time which states that the ordinal structure of time is not needed even in our classical theories, especially in general relativity. In this article I analyze this position and the arguments supporting it, and I argue that there are some serious concerns with some of the radical relationalists' arguments which make it an unattractive position. In this sense, I conclude that the chrono-ordinal structures of our theories play important theoretical and explanatory roles and that they can be taken to be part of the empirical content of our theories. 
\end{abstract}


\section{Introduction}

The search for a theory of quantum gravity invites us to consider to what extent space and time could be quantum and some approaches claim that time could even be absent from our theories. This represents a huge departure from the way we normally understand our physical theories, and hence it creates some resistance to accept these approaches to quantum gravity\footnote{See \cite{Esfeld2019,MozotaFrauca2023, mozota_frauca_problem_2024, mozota_frauca_quantum_2025}.}. Part of the quantum gravity community has reacted to this, arguing that even for theories like Newtonian mechanics or general relativity we can eliminate our concept of time or reduce it to a minimum, and hence that the conceptual gap between our classical theories and quantum gravity isn't as big as their critics have argued\footnote{See \citep{Rovelli2011}.}. In this sense, they embrace a new metaphysics of time, which I, following other authors\footnote{See the characterization of radical relationalism by Thebault in \citep{Thebault2012,Thebault2021}.}, call radical relationalism, although it is rather an eliminativist position. In this article I will oppose this position and argue that time, at least in its ordinal role, is indispensable for our classical theories.

The issues discussed in this article are important from the point of view of both theoretical physics and philosophy of physics. My article is the first to characterize in some detail and oppose the metaphysics of time that lies behind some of the arguments from the quantum gravity community. In particular, by arguing that time is indispensable for our understanding of classical systems, I claim that the conceptual gap between putatively timeless theories of quantum gravity and classical theories is bigger than what is depicted by the quantum gravity community. In this sense, I argue that the defenders of the viability of some approaches to quantum gravity, both in the physics \citep{Rovelli2004,Rovelli2011} and philosophy literature \citep{Huggett2013a,Lam2018-LAMSIA-2,Lam2020} face a great challenge that shouldn't be addressed by endorsing an unappealing reading of our classical theories.

The structure of this article will be the following. I will start in section \ref{sect_radical_relationalism} by introducing radical relationalism and the existing motivations for holding such a position. Then, I  will present my arguments for resisting it. In section \ref{sect_resisting} I discuss radical relationalism from a more conceptual point of view by studying a system usually described by a Newtonian model and I observe that while there is room for some relationalism about time, the ordinal structure of time seems indispensable for explaining and predicting the behavior of this kind of system, and hence that one can resist radical relationalism. Then, in section \ref{sect_gr} I will generalize my conclusions to the case of general relativity. That is, I will argue that in general relativity there is also a chrono-ordinal structure that is an essential part of its models, its predictions, and its explanations. For this reason, radical relationalism is an unappealing position, as it abandons a fundamental ingredient of the theory, jeopardizing our understanding, explanation, and predictions for a wide variety of phenomena. Finally, in section \ref{sect_implications} I comment on the implications that this can have for the philosophy and foundations of quantum gravity.

\section{Radical relationalism}\label{sect_radical_relationalism}

Radical relationalism appears as an eliminativist position about time in classical systems that part of the quantum gravity community is embracing. In this section I will outline more clearly this position and the arguments supporting it.

Relationalist positions in the philosophy of time or spacetime are usually based on the intuition that our theories postulate too many unobservable spatiotemporal structures, and they propose a meager ontology in which these structures are not needed but which is still able to account for the phenomena that our theories predict. 

The classical example of a relationalist argument is Leibniz's argument \citep{Leibniz1715} against Newton's absolute space. If we imagine a world that is a copy of the actual world but in which every body is 3 feet to the right (according to some arbitrary reference axis), this would be observationally indistinguishable from the actual world. The reason for this, the argument goes, is that in a Newtonian world we do not observe the distances with respect to absolute space, but just the relative distances between bodies. As absolute space is not empirically relevant, Leibniz proposes eliminating it and postulating an ontology in which there are bodies, and spacial relations in between them, but no absolute space. 

The paradigmatic example of a relationalist argument in contemporary spacetime metaphysics is the hole argument \citep{Earman1987}, which follows a similar structure: it takes two general relativistic models related by a diffeomorphism to be observationally indistinguishable, and then concludes that the two models shouldn't be taken to represent different metaphysical possibilities. There are many philosophical subtleties involved with this argument and its extent, but for the discussion in this article it will be enough to give this broad characterization\footnote{The hole argument is also closely related to matters such as determinism. I refer the reader to \citep{Norton2019} and references therein for careful philosophical discussions of the argument. See also \citep{Hoefer1996,pooley_hole_2022} for responses to this argument.}.

Similarly, relationalist arguments can be rehearsed against absolute time, absolute durations, or absolute lengths. In this sense, relationalist positions postulate that some or all of these spatiotemporal relations are just relative to physical bodies or processes and tend to eliminate absolute, unobservable structures from the ontology and keep just the bare minimum to have an interpretation of the formalism which is able to account for the empirical data of our spacetime theories.

Despite this eliminative agenda, there is a minimal spacetime structure that relationalist positions tend to keep. This is the chrono-ordinal or causal structure of spacetime. That is, for most relationalists it is still true that events can be located with respect to one another. For instance, the claim that the instant in which I am writing this happens between the instant in which I first thought about writing this article and the instant in which you are reading it is considered a true and meaningful fact by the majority of relationalists. Indeed, Leibniz speaks about space as an `order of coexistences' and time as an `order of successions' \citep{Leibniz1715}\footnote{I am grateful to an anonymous reviewer for raising this point.}. In this sense, we can see how ordinal structures\footnote{Let me clarify here that by ordinal structure I refer to any structure that defines an in-betweeness relation between instants or events, but that this does not necessarily mean that this structure is directed. In other words, not every ordinal structure will introduce a distinction between past and future. Indeed, different views of time disagree on whether such a distinction is necessary. In contemporary physics, the general trend is to reject the distinction based on the symmetries of the laws of motion of our physical theories. This is in contrast with authors like Leibniz, who defended this asymmetry. See \cite{lopez_relational_2025} for a discussion of this point.} play an important role in relationalist positions.

However, researchers in the quantum gravity community \citep{Rovelli2004,Rovelli2011} and philosophers of physics studying the foundations of quantum gravity seem to embrace a stronger version of relationalism, radical relationalism, which eliminates temporal structure altogether, including the chrono-ordinal and causal structure of spacetime. Part of the motivation for adopting this view has to do with the timeless structures found in some approaches to quantum gravity, but it is a view that aims to apply and be well-motivated in the context of general relativity. In this sense, in this article I will study the motivations for radical relationalism in the context of classical spacetime theories and argue against it.

In ordinary physics there is a distinction between the variables and structures describing space and time or spacetime and the variables describing what is there in space or what is happening in spacetime. That is, there is a clear conceptual distinction between the part of our theories that answers what questions and the part that answers when or where questions. Radical relationalists claim that one of the lessons of relativity theory is that there is no privileged variable in our theories and that we should put all the variables, including time variables or variables describing spacetime, `on the same footing'. Moreover, in some models like in general relativity there may not be, according to radical relationalists, any time variable, and in this case one can still make sense of the model, as from their point of view they are not necessary.

To clarify how they claim this is possible, let me introduce the work of Carlo Rovelli as the most influential radical relationalist. Rovelli \cite{Rovelli2002a,Rovelli2004,Rovelli2011} argues that physics is about correlations between physical quantities we can observe and measure, which he calls partial observables. In this sense, physics would be about how some measurements correlate with some others and not about how things develop in spacetime. For instance, when we consider a Newtonian model of a system of three particles it encodes the correlations between the measurable positions $x,y,z$ and the measurable time $t$. But it is important that, according to Rovelli, time is not special, and the model predicts correlations of the form $x(t)$, i.e., the position of a particle as a function of time, and also $x(y)$, the position of a particle as a function of the position of some other particle. Rovelli argues that if we many times describe physics in terms of time variables it is just because of convenience and not because they capture something fundamentally different from what is represented by other variables and structures in the formalism. Indeed, for Rovelli what $t$ is ultimately representing is the state of some clock we use for keeping track of time.

The departure is even bigger when Rovelli considers models like general relativity or theories with a reparametrization invariance. An example that Rovelli discusses often can be found in \citep{Colosi2003}. This is a model containing two variables $x$ and $y$, and Rovelli interprets it not as describing how two degrees of freedom evolve in time or some chrono-ordinal structure, but just as giving us the set of possible outcomes from our measurements of both degrees of freedom. That is, the model gives us a set of possible pairs $x,y$ but there is no order in this set. While in a standard interpretation of this kind of model there would be a fact about whether a pair $x_1,y_1$ happens in between the pairs $x_2,y_2$ and $x_3,y_3$, for Rovelli this is not part of the model. Again, it is just correlations between physically measurable quantities that matter for him.

In his essay \textit{Forget time} \citep{Rovelli2011} Rovelli makes it explicit that his aim is a formulation of classical physics in which time plays no special role. His section on the thermal time hypothesis makes it clear that for him spatiotemporal structure is not something fundamentally different from what is represented by other variables in our physical theories, but for him it is something to be associated with thermodynamics. In this sense, it should be clear that Rovelli is advocating for something quite radical in which there is no fundamental distinction between when and what questions and in which the chrono-ordinal structure of time plays no important role.

At this point, let me clarify that radical relationalism may be seen not as a relationalist view of time, but rather as an atemporal view of reality. By starting with some relationalist intuitions about evolution, what advocates of radical relationalism end up proposing is a view in which there is no time left. They would claim that some temporal intuitions such as relational evolution or thermodynamical senses of time are still there, but the key point is that the more basic intuition about time, that of defining sequences of events, is missing from this perspective. In this sense, I will take radical relationalism to be an eliminativist view of time, although my arguments in this article will apply equally to characterizations of the view which claim that there are some temporal features (not temporal ordering) preserved. 

Finally, let me mention that radical relationalists have two sources of motivation and argument for their position, besides the motivation from quantum gravity. The first one is related to the motivation of standard relationalism and it is the position that the ontology should be as close to our observations as possible and that we should avoid introducing unobserved structures which are not required to explain the empirical content of our theories. This relationalist motivation leads radical relationalists to reject the chrono-ordinal structure of spacetime, in the same way that motivated moderate relationalists to reject absolute space. The second motivation for radical relationalists has to do with the interpretation of the reparametrization invariance of general relativity as a gauge symmetry.

It is in the context of gauge theories like electromagnetism in which the notion of observable was introduced. In gauge theories we have multiple equivalent representations of the same physics. For instance, in the case of electromagnetism, the physically meaningful electromagnetic field $F_{\mu\nu}$ is represented by the 4-potential $A_{\mu}$, and this representation is not unique, i.e., there are multiple $A_{\mu}$ which represent the same field. For theories like electromagnetism, there is a formal definition of `observable' which allows us to tell apart which part of our model is representation-dependent and which is representing the genuine, physical content of the model. In the case of electromagnetism, this definition leads to the expected conclusion that $F_{\mu\nu}$ is observable, while $A_{\mu}$ isn't, just as one would expect.

Radical relationalists apply this criterion of observability to theories like general relativity to conclude that the physical content of the theory has nothing to do with points in a manifold but lies precisely in the `observables' that satisfy the formal definition that applies in the case of electromagnetism. These observables would be nothing but the correlations between partial observables introduced by Rovelli. Authors like Earman \citep{Earman2002}, have taken the application of the observability criterion to theories like general relativity to imply that the interpretation of the theory has to be radically different: while in other spacetime theories one can have a standard relationalist or substantivalist metaphysics of time, in general relativity one has to abandon it and embrace a new alternative. Authors like Rovelli propose radical relationalism, the view that the physics of general relativity is contained in the correlations between physical variables (not necessarily including time variables or other spatiotemporal structures), as the correct way of interpreting the result of applying the observability criterion.

Some authors \citep{Kuchar1993,Maudlin2002,Pitts2014-PITCIH,Gryb2016,Pitts2017,MozotaFrauca2023, mozota_frauca_gps_2024}  have raised serious objections to the application of this criterion to the case of general relativity and to the conclusions reached from them. The argument in this paper will be more focused on the conceptual part than on the technical part and I won't be giving an explicit comparison with the case of electromagnetism. The core of the more technical objections has to do with the different nature of the symmetry transformations between two physically equivalent models in the cases of electromagnetism and general relativity. While in electromagnetism these transformations affect the way in which the electromagnetic field at a point is represented, in the case of general relativity the transformation affects the representation of spacetime points themselves. Given this substantial difference, one cannot treat both types of transformations in the same way and jump to metaphysical conclusions.

\begin{figure}[ht]
\centering
\includegraphics[width=0.9\textwidth]{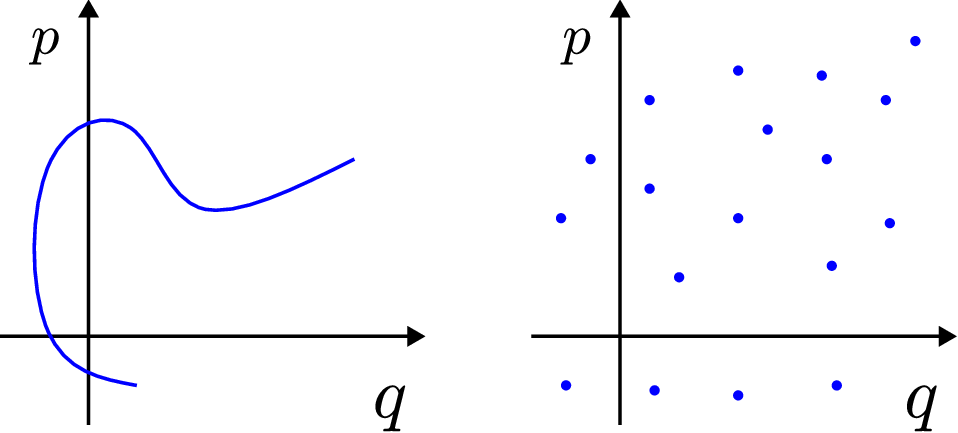}
\caption{\label{figure_phasespace} On the left-hand side, a trajectory in phase space respecting the one-dimensional chrono-ordinal structure of our physical theories. On the right-hand side, a set of points with no ordinal structure. Radical relationalism is compatible with the right-hand side picture, but our best physical theories are represented by the left-hand side.}
\end{figure}

To insist, radical relationalism is the view that the physical content of our theories lies just in the correlations between the physical variables they predict. Spatiotemporal structures wouldn't be special structures in our variables or they could even be absent (see an example in figure \ref{figure_phasespace}). In this sense, this version of relationalism is more radical than other versions. While in substantivalist views and milder relationalist views there is a fundamental distinction between spacetime and the things that happen in spacetime, radical relationalism eliminates this distinction.

\section{Resisting radical relationalism}\label{sect_resisting}

I will now argue against radical relationalism, i.e., I will argue that the best way of understanding our physical theories and the world is as containing two fundamentally different ingredients: spacetime or spatiotemporal relations and matter, bodies, physical entities, physical processes, or whatever we consider that there is in spacetime. In this way, I argue that the chrono-ordinal structure of spacetime is an ineliminable part of our physical theories and of the world. In other words, I will argue that the best ways we have to describe and explain a wide variety of phenomena rely on assuming that the world has some chrono-ordinal structure, and hence that it cannot be eliminated from our physics and metaphysics without paying a high price. While I take it that it is a widely held position, my account is the first to address the radical relationalist arguments in the quantum gravity literature.

The discussion is divided into two parts. First, in this section I will give a general argument against the radical relationalist position from a conceptual point of view. I will analyze a toy model and argue that the chrono-ordinal structure is an indispensable part of the system it describes. Second, in section \ref{sect_gr} I will address some of the specific arguments that are based on the structure of general relativity.


Let me discuss a model of pre-relativistic physics and the way radical relationalism would apply to it. The model is a solar system model in which we describe the positions of $n$ planets $x_1, x_2, x_3 ... x_n$ as they orbit around the Sun, with position $x_S$ . On top of this, we can have $n$ additional variables $\alpha_1, \alpha_2,... \alpha_n$ measuring the angles of rotation of each planet with respect to its rotation axis. By applying Newton's equations of motion we are able to obtain full predictions for the state of the solar system at any time, i.e., the set of functions $x_S(t), x_1(t), \alpha_1(t), x_2(t), \alpha_2(t),...$, where $t$ is Newtonian absolute time.

Moderate relationalists claim that absolute time is not necessary in this picture. Leibnitz argued that if we shifted time by an arbitrary quantity $t\rightarrow t+t_0$ the empirical content of the model would be the same. Along the same lines, authors like Mach propose a more general transformation in which we replace Newtonian time with any arbitrary monotonic function, i.e., $t\rightarrow \tau(t)$. Even for this general transformation, a case can be made that the empirical content is preserved.

For instance, focus just on the pair of functions $x_1(t), \alpha_1 (t)$ obtained by solving the Newtonian equations of motion and the functions $x_1(\tau),\alpha_1(\tau)$ obtained by transforming $t$ into an arbitrary monotonic parameter $\tau$. One can argue that they make the same predictions: they both describe a planet orbiting around the Sun and spinning around its axis. Furthermore, if we focus on one given `year' for this planet, i.e., in its trajectory around the Sun once, both descriptions  $x_1(t), \alpha_1 (t)$ and $x_1(\tau),\alpha_1(\tau)$ agree in how many `days', as defined by complete spins of this planet around its axis, this trip takes. That is, in both models we can say that one `year' for this planet lasts $N$ `days'. For the moderate relationalist this is a meaningful prediction and closely related to the way we operationally define durations. 

Now, the model with the Newtonian time can make one more prediction. It can also say that one `year' lasts a given absolute duration, say $Y$ seconds, and it can relate it with what other systems would do in this period. That is, this absolute duration allows us to relate the functions $x_1(t),\alpha_1(t)$ with other systems, even if we don't explicitly consider them. For instance, if we know that the rotation period of the Earth is 24 hours, we can relate $Y$ with the rotation of the Earth to deduce how many Earth-days will take planet 1 to complete a turn around the Sun.

The relationalist however can claim that this prediction is also included in the model using the $\tau$ parameter. In this case, what one should do to extract this prediction is to study not just the two functions $x_1(\tau),\alpha_1(\tau)$ but also the function $\alpha_ 3(\tau)$ representing the rotation angle of the Earth. By doing so, one can relate the trajectory of $x_1$ with the rotation movement of the earth and deduce how many Earth-days elapsed in one `year' of the first planet. 

In this way, for the moderate relationalist, the physical content encoded in the Newtonian time parameter is just to relate the processes described by the model with some other processes. Once our model includes those other processes, there really is no need to have absolute time, as the relative durations are described by the model even using an arbitrary temporal parametrization. Of course, for all practical purposes, it is far more convenient to have absolute time in our models than to include every physical system we may want to use to perform duration measurements. That is, if we want to relate the solar system model with the time measurements of my watch, we would definitely use Newtonian time instead of having a model including the mechanism of my watch.

The difference between the relationalist and substantivalist or defender of absolute durations is what to make of this convenient parameter. While for the relationalist it is just a powerful idealization or convention, for the substantivalist the fact that it is so powerful and present in our theories signals that it is not just a mere convention but a structure in the world. Another way in which defenders of absolute durations argue for it is that duration allows relating processes not just to other processes in the world, but with counterfactual processes. For instance, in the solar system model absolute duration can be associated with what clocks in some of the planets would measure, even if no such a clock existed.

Now we can turn to what the radical relationalist would say about this example. While the moderate relationalist stripped down the original functions $x_1 (t), \alpha _1 (t)$ to the arbitrarily parametrized $x_1(\tau), \alpha _1(\tau)$ which still maintained the chrono-ordinal structure of time, the radical relationalist would even eliminate this structure and keep just the correlations between $x_1$ and $\alpha_1$. Formally, we could say while the predictions for the moderate relationalist are in the equivalence class of functions $x_1(\tau), \alpha _1(\tau)$ under temporal reparametrizations, for the radical relationalist the predictions are not the equivalence classes but the (unordered) set of pairs of values $(x_1,\alpha_1)$, independently of the order in which they obtain in $x_1(\tau), \alpha _1(\tau)$. While for the moderate relationalist there is a fact about whether a pair $(x, \alpha)$ occurs in between two other pairs $(x',\alpha')$ and $(x'',\alpha '')$, for the radical relationalist there wouldn't be such a fact.

If we take the values of $x_1$ along a given turn around the Sun each time the angle $\alpha_1$ is $0$, i.e., at one `day' intervals, we obtain a sequence of positions: $X_0,X_1,X_2...$. While the moderate relationalist takes it that this order is a meaningful part of the model, the radical relationalist doesn't. The sequence $X_0, X_2, X_1, X_3...$ would be considered physically wrong by the standard understandings of Newtonian models, but for the radical relationalist it would be perfectly fine, as for them it is just the set of values and not the order at which they happen that is physically meaningful.

\begin{figure}[ht]
\centering
\includegraphics[width=0.9\textwidth]{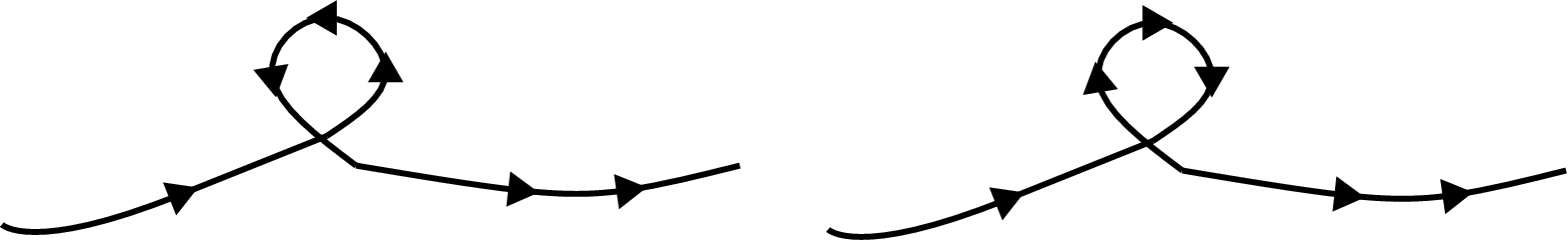}
\caption{\label{figure_loop} Two possible ways of ordering a trajectory. While the standard view would take only the first one to be physical, radical relationalism would take both to be equivalent.}
\end{figure}

A similar example is the case in which we have a trajectory that intersects with itself like in figure \ref{figure_loop}. This could be the case of a planet that is moving in space, turns around some massive body, and then continues with its trajectory. This trajectory would be different from a trajectory that agreed on the points but which had the loop inverted: if in the original trajectory the loop was traveled anticlockwise and the planet was spinning just in the same direction as before, during, and after the loop, in the alternative one the loop is traveled clockwise and the spinning of the planet is in the opposite direction than before or after the loop. Clearly, the intuition that we would have for this second trajectory is that it is unphysical: planets do not abruptly change the direction of their motion and spin. For the radical relationalist both trajectories would be equally fine, as the radical relationalist denies there is something physically meaningful in the order we assign to physical correlations. 

I see two ways in which the radical relationalist could respond to this kind of problem. The first one is to appeal to relationalism and claim that the apparently natural and right temporal order is encoded in correlations with external systems that we use as clocks. In the example of the planet we can use the the spin of the planet to pick a privileged ordering of instants in the trajectory, that is, of the two possibilities described above, we conventionally pick the one in which the planet spin does not abruptly change direction. Alternatively, we can use correlations with other systems, like a free body moving in a straight line. If we conventionally (according to the radical relationalist) pick the free body to define an order, we can choose the most natural one and not the seemingly unphysical. But again, for the radical relationalist this is just a matter of convention: out there in the world there is no chronological order, there are just correlations between physical quantities.

We can see how this strategy seems a bit strange. While in the case of duration it (arguably) makes some sense to claim that duration is relational and that if we include all the degrees of freedom of the universe in our model we do not need to have an absolute duration, the same claim seems much less justified for the case of the ordinal structure of time. That is, the chrono-ordinal structure of our models plays a fundamental role in our physical theories. Our best physical theories are written in terms of systems of partial differential equations that relate how physical magnitudes vary infinitesimally and continuously in spacetime. In this sense, we rely on the chrono-ordinal structure of time to explain physical phenomena. Similarly, concepts like locality or causality, which play important explanatory roles in our understanding of physics, also rely on this temporal structure. If the radical relationalist gets rid of time, physics as we understand it and its explanatory power seems to be lost.

Along these lines, physical magnitudes that we commonly take as physical, meaningful, and even predictive when we have knowledge about the laws of nature, also rely on the temporal structure of the world. These quantities are velocities and other rates of change. The radical relationalist is forced to either abandon velocities as physical quantities or rethink a definition for them other than the standard ones relying on temporal derivatives.

The second line of answer of the radical relationalist could be to claim that the appearance of a temporally ordered world is not something empirical or for physics to directly explain, but associated with our experience and consciousness. This kind of argument is similar to the way some authors have argued that our impression that there is an arrow of time, a difference between past, present, and future does not capture any fundamental asymmetry in the structure of the world or laws of nature, but it is just `accidental' and it just aligns with the direction of local entropy increase. This is to say that if we had a brain (or some other conscious object) in a region of spacetime in which entropy decreased with time, it would experience time `backwards', i.e., in the opposite direction to the brains in the regions of the universe in which entropy increases. Now one could argue that in the same way that our experience of a temporal asymmetry could be explained without a fundamental temporal asymmetry in the world, our experience of temporal order could be explained without time in the world. The claim there would be that in the set of correlations there are the correlations corresponding to a brain interacting with its environment, then the brain would have the experience of a temporally ordered sequence.

This kind of argument goes beyond philosophy of physics and of science and enters into debates in the philosophy of mind such as the relationship between the physical and the mental. However, without entering into the details of the mind-body problem, we can see how radical relationalism is also radical and unappealing regarding this connection. It proposes an ontology which makes the connection between our experience and what is there in the world more remote. Ironically, we have seen how one of the motivations was an empiricist one that wanted to keep physics as close to what we observe as possible. Of course, if there is some good reason for adopting a metaphysics that does not seem to fit well with our experience, this wouldn't necessarily be a major problem, but for the case of radical relationalism I will argue that we are lacking such a good reason.

Finally, let me turn to the more technical part of the discussion. Moderate relationalists can give a formal description of Newtonian mechanics which reflects their philosophical position. Standard Newtonian models can be defined by means of the following action principle:
\begin{equation}
S[q]=\int dt (T-V) \, ,
\end{equation}
where $q$ represents the configuration space variables describing our system and $T$ and $V$ the kinetic and potential energy of the system. In our example $q$ would correspond to the set of $x_i, \alpha_i$, $T$ would be the sum of kinetic energies of each body (including rotational kinetic energy), and $V$ would describe the gravitational potential energy. By extremizing this action one arrives at the standard Newtonian equations of motion and solutions $q(t)$.

Now, the moderate relationalist can offer a different formalism\footnote{In particular, Julian Barbour has extensively used this formalism to defend his philosophical views \citep{Barbour1994,Barbour2011}.}. This is defined by means of the Jacobi action:
\begin{equation}
S[q]=\int d\tau \sqrt{T(E-V)} \, .
\end{equation}
$T$ and $V$ have the same form as in the standard action, and $E$ is a constant that can be interpreted as the total energy of the system. Now we do not have Newtonian time $t$ in the action, but an arbitrary time parameter $\tau$. Velocities (linear, angular, or of any kind that appear in the model) are also defined with respect to $\tau$, and hence despite the formal similarity $T$ is not exactly the kinetic energy of the system. When we extremize the action to find the equations of motion, what we find is different from the set of Newtonian equations of motion. One finds one fewer independent equation than variables, which means that instead of having a unique solution for a set of initial conditions, one has an infinite set of solutions $q(\tau)$. However, this is not as bad as it sounds, as all the solutions are equivalent according to the relationalist standards. The reason for this is that all the solutions (for equal initial conditions) are related by a temporal reparametrization, i.e., if we have a solution $q(\tau)$ we can obtain another one by transforming $\tau \rightarrow f(\tau)$, where $f$ is any monotonic function. In other words, different solutions are equivalent in that they describe the same sequence of configurations of the system, even if they differ in the temporal label $\tau$ they ascribe to the instants in this sequence.

This formalism captures well the moderate relationalist intuitions. It is a formalism in which the metric aspect of time is not explicitly there, while the chrono-ordinal structure is there and is the same for every equivalent solution. The moderate relationalist can argue that this formalism represents what they take to be the empirical content of Newtonian physics (relative durations and order) with no need to introduce absolute duration.

Radical relationalists, however, also take this model to exemplify their intuitions. Their argument is based on the analysis of the reparametrization invariance of the model as a gauge symmetry. This model shares with gauge theories like electromagnetism that it defines equivalence classes of solutions. In the case of the Jacobi action we have the equivalence class of all the possible ways to parametrize a sequence of configurations of the system, while in electromagnetism we have the equivalence class of all the 4-potentials that represent the same electromagnetic field. The radical relationalist is right in claiming that there is a similarity here and arguing that we shouldn't take the details of any member of the equivalence class to be physically significant if they are not present in any other member. However, the way in which they propose to find the shared physical content is by finding the `observables', which as I have mentioned above I take to be fundamentally mistaken for the case of models with a reparametrization invariant.

Take the case of the Jacobi action for our solar system model. The equations of motion give us an equivalence class of solutions $x_i (\tau), \alpha_i (\tau)$ and the radical relationalist claims that what is common in this equivalence class is the correlations between the different magnitudes. They are right that all the solutions in the equivalence class agree in predictions like what will the position of the Earth be when Mars is at some other position, but they are wrong in claiming that that's all that solutions in the equivalence class have in common. As I have argued, claims about relative duration and order are also part of the equivalence class, and correlations fail to capture them. For this reason, the analysis of the symmetry of the Jacobi action does not support the radical relationalist conclusion. Chrono-ordinal structure is part of this kind of model, contrary to their reading of the model.

In this sense, while the moderate relationalist can provide a formalism that reflects their intuitions and which is arguably empirically equivalent to Newtonian physics, the radical relationalist hasn't provided so far an alternative formulation of physics in which one could argue that the ordinal structure is missing while maintaining the same predictive and explanatory power. Indeed, in all the discussions of Newtonian systems and models by authors like Rovelli, one finds the one-dimensional chrono-ordinal structure in some way or another even if it is claimed that only correlations matter and that this ordinal structure is just accessory. That is, one still finds models in which one has action principles defining one-dimensional trajectories in configuration or phase space with equations of motion relating one configuration with the following.

Radical relationalism faces then the following problem: if it is only correlations that matter, why is it that their models inevitably end up invoking one-dimensional ordinal structures? In other words, radical relationalism is compatible with theories in which one defines a cloud of correlations (figure \ref{figure_phasespace}), or phase-space regions of dimensionalities other than one, but it is still choosing models that define, even if implicitly, a one-dimensional ordinal structure. While the radical relationalist can claim that this is done in order to represent physics such as we understand it today, this seems to be an argument that supports the conclusion that our best physics suggests that there is a one-dimensional chrono-ordinal structure in the world.

\section{General relativity}\label{sect_gr}

The discussion so far has focused on the conceptual issues in a pre-relativistic setting. However, the discussion can be easily generalized to the case of general relativity.

In the case of general relativity we cannot speak about time and space anymore but we have to deal with spacetimes. The structures of our models are therefore far more complicated. Despite this, our models still have a clear chrono-ordinal structure. The causal structure of spacetime (and a choice of orientation) allows us to define the past and future of any point. That is, relative to any point in spacetime, we can classify the rest of the points into three sets: the set of events in its future, the set of events in its past, and a third set of points that are neither in its future nor in its past. In this sense, we can still make claims such as that relative to the day I am writing this article, the event of you reading it is in the future, and the event of me first thinking about writing it is in its past.

Consider the example of a solar system model. In the general relativistic case, the model will give us a family of functions $x_S(\tau), x_1(\tau), \alpha_1(\tau), x_2(\tau), \alpha_2(\tau),...$, for some arbitrary time coordinate $\tau$, together with a metric tensor $g_{\mu\nu}$. In this model there is still an order relation encoded by $\tau$ and the metric. When we consider a single planet, we can still claim that its position and angle at a time $\tau_1$ comes in between its positions and angles at times $\tau_0$ and $\tau_2$. However, now we cannot say that the position of planet 1 at coordinate time $\tau_1$ and planet 2 at the same coordinate time happen simultaneously, as in relativistic theories simultaneity does not play any role. What one can say (by studying the structure of spacetime) is that relative to a coordinate time and planet, say Earth at time $\tau_0$, one can divide the trajectories of the rest of the planets into three parts: the part lying on the past of Earth at $\tau_0$, the part lying on its future, and a third part which is neither in the past nor the future. 

While in Newtonian models reading the chrono-ordinal structure of time was immediate, even in arbitrarily parametrized models, in general relativistic models this is not so straightforward. For models like the solar system model just discussed, some time coordinates may encode some order relations, but not all of them. To determine whether two spacetime points are one in the future of the other in general one needs to study the metric tensor of spacetime, which is the piece of the formalism that encodes the chrono-ordinal structure of spacetime.

Just as in the case of arbitrarily parametrized Newtonian models, in the case of general relativity we also have equivalence classes of models which represent the same spacetime but for a different choice of coordinates. This, however, does not affect the chrono-ordinal structure of spacetime: if an event is in the future of another one according to one model in the equivalence class, it also is according to any other model in the class. The differences with the Newtonian models discussed before are that now the equivalence classes are much bigger, as we are allowing for arbitrary changes of coordinates in four dimensions, and that the chrono-ordinal structure is now a partial order structure, reflecting the fact that not every pair of points is ordered.

In this sense, the claims and arguments in the previous section can be adapted to the general relativistic case to argue that the order structure of spacetime is taken as a prediction of our models, that it plays an important role in our theories and that therefore radical relationalism is not an attractive position.

For instance, in the previous section I argued that radical relationalism is in trouble at the time of accounting for relative durations. When there is an order structure it makes sense to claim that one year for one planet lasts $N$ days: we can extract this information by counting how many times the planet spins around its axis in between the initial and final moments which define the year. If the chrono-ordinal structure is missing, we are lacking this in-betweenness relation and relative durations are not easily recovered. In the case of general relativity, the same argument for the case of the planet can be applied in exactly the same way, as the planet follows an ordered trajectory in spacetime and one can use this structure to count the number of revolutions in between two points of the trajectory. For relative durations between the motion of different bodies, the technical details of how to define them are more complicated and involve a (small) degree of conventionality, but one can still make use of general relativistic models to make claims about them. That is, in general relativity we can still make claims like that a Venus-year lasts 225 Earth-days\footnote{This kind of claim can be justified from the perspective of an observer on Earth, from the perspective of an observer on Venus, by some other perspective, or by introducing a foliation of spacetime. The order of magnitude of the difference between the different ways of defining the duration is bounded above by the time it takes for light to travel between Earth and Venus (a few minutes at most, which is a small margin when defining durations of the order of years).}, and radical relationalism fails to account for this kind of facts.

Just as in the case of Newtonian physics, we take facts about order to be part of the empirical content of the theory. For instance, a general relativistic model may predict the following events: the merge of two black holes, the relaxation of the resulting one, the emission of a gravitational wave, and its reception and detection by LIGO. General relativity does not only predict these events as (spatiotemporally) local states of affairs or correlations, but as an ordered sequence. Moreover, the chrono-ordinal structure plays a major role in making these predictions, and it is generally taken to play an explanatory role. The causal structure of general relativity is captured by its equations of motion and the equations of motion describing matter, the electromagnetic field, or any other ingredient that we may have in our models. From this point of view, giving up the chrono-ordinal structure of general relativity implies giving up a fundamental ingredient of the theory.

As commented in section \ref{sect_radical_relationalism}, one of the arguments by radical relationalists is the analysis of the diffeomorphism invariance of general relativity as a gauge symmetry. This argument can also be blocked in the same way I argued in section \ref{sect_resisting} that can be blocked for the case of the Jacobi action. That is, while it is true that one should take the physical content of a model to be independent of the representational choices one makes for describing it, it is false that in the case of reparametrization invariant theories this makes the chrono-ordinal structure of spacetime unphysical. In general relativity we have a family of diffeomorphism-related models for describing the same spacetime, but they agree in more than just the correlations between observables: they agree in their temporal structure. For instance, we may have two models describing our solar system and including the events of me writing these lines and of you reading them. These models differ in many representational details such as the points of the manifold or coordinates ascribed to these physical events, but they agree that my writing happens before your reading. In this sense, analyses of the gauge structure of general relativity that conclude that the temporal structure of spacetime is not part of the theory are missing an important ingredient of our models.

The technical complications associated with dealing with the symmetries of the formalism of general relativity may lead us astray and embrace a radical version of relationalism that denies the explanatory, theoretical, and empirical importance of the temporal structure of the theory. As I have argued, I find that there is no convincing argument, technical or conceptual, based on our classical theory of general relativity or on the phenomena it explains and predicts, to endorse such relationalism.

\section{Radical relationalism and quantum gravity}\label{sect_implications}

In this section I will make a remark about the relationship between radical relationalism and some approaches to quantum gravity.

One of the motivations behind radical relationalism is to have a view of classical general relativity which is compatible with some present or future theory of quantum gravity. It is expected that the formulation of a quantum theory of gravity will imply important departures in the way we conceptualize space and time, and in this sense we can see radical relationalism about time in general relativity as a position that aims to facilitate the connection between general relativity and some theory of quantum gravity in which temporal structures are meager. However, the arguments in this article support the conclusion that the temporal structures of our classical theories are not so easily disposable.

This connects with the philosophical discussion regarding the possibility and sensibility of a timeless theory of quantum gravity. Different philosophers \citep{Huggett2013a,Lam2018-LAMSIA-2,Lam2020} have argued that as long as the theory of quantum gravity is able to recover spacetime or spacetime functions as some approximation in some regime, then the theory of quantum gravity could be fundamentally timeless. This is sometimes phrased in terms of two theories with a conceptual gap between both of them. Radical relationalism can be seen as a proposal for a way of narrowing this gap: by denying that chrono-ordinal structure is a physically significant part of our models or that it is necessary for understanding our classical theories, it aims to make general relativity conceptually more similar to some allegedly timeless theory of quantum gravity.

The analysis in this paper goes against this proposal, as it supports the conclusion that the chrono-ordinal structure of general relativity is a fundamental ingredient of the theory. According to my position, the challenge for the quantum gravity theorist or philosopher of quantum gravity who defends these models is greater than what is usually discussed\footnote{See some complementary arguments in \citep{MozotaFrauca2023}.}. The positive reading of my argument is that it points to the direction spacetime functionalists should go if their position is to have any traction: instead of denying the role that chrono-ordinal structure plays in our theories, it should aim to explain how it could be recovered or approximated.

In any case, it should be clear that taking quantum gravity as an argument for radical relationalism is, to say the least, an unattractive position, as these theories and models aren't well-established and there are many conceptual challenges that these theories have to face. In this sense, from my point of view, theories of quantum gravity, as they stand right now, do not represent a solid foundation upon which to build a philosophical position.

%
%
%
%
%


\section{Conclusion}\label{sect_conclusions}

In this article I have argued against radical relationalism, the philosophical position which takes the chrono-ordinal structure of our theories to be disposable, especially in the case of general relativity. I have argued that while some other forms of relationalism are more reasonable, radical relationalism eliminates a structure that we have good reasons to believe that forms part of the empirical content of our models and that plays important theoretical and explanatory roles. Moreover, some of the technical arguments employed by the radical relationalist, namely those based on the gauge structure of the theory, can be challenged, as the chrono-ordinal structure of general relativity is invariant under diffeomorphisms.

Finally, I have argued that this analysis of radical relationalism weakens their position regarding some approaches to quantum gravity, as I have argued that the gap between some approaches to quantum gravity and classical theories is bigger than what they argue to be.

According to my perspective on spacetime in classical theories, chrono-ordinal structures encode empirical facts and are a fundamental ingredient for the explanatory power of these theories. In this sense, I think that, if there is a way of making sense of timeless theories of quantum gravity, it is not by eliminating time from our classical theories, but by paying more attention to the way in which it could be recovered. In other words, I believe that for `timeless' approaches to quantum gravity to be appealing, more work should be done to show how chrono-ordinal structures could be encoded or approximated in these approaches instead of denying their physical significance.

\section*{Acknowledgements}
I want to thank the Proteus group, Carl Hoefer, and the audiences at Buenos Aires and Belgrade for their comments and discussions. 

This research is part of the Proteus project that has received funding from the European Research Council (ERC) under the  Horizon 2020 research and innovation programme (Grant agreement No. 758145) and of the project CHRONOS (PID2019-108762GB-I00) of the Spanish Ministry of Science and Innovation.

%


\printbibliography

\end{document}